\documentclass[floatfix,twocolumn,showpacs,showkeys,preprintnumbers,amsmath,amssymb]{revtex4}
\usepackage{graphicx}% Include figure files
\usepackage{dcolumn}% Align table columns on decimal point
\usepackage{bm}% bold math
\begin{document}

%\preprint{APS/123-QED}

\title{
\boldmath $K^-\,^3\mathrm{He}$ and $K^+ K^-$ interactions
in the $pd\to{}^3\mathrm{He}\,{K^+K^-} reaction$
}%

\author{V.Yu.~Grishina}
\affiliation{%
Institute for Nuclear Research, $\mathit{60^{th}}$ October
Anniversary Prospect 7A, 117312 Moscow, Russia
}%
\author{M.~B\"uscher}
 \email{m.buescher@fz-juelich.de}
\affiliation{%
Institut f\"ur Kernphysik, Forschungszentrum J\"ulich,
D-52425 J\"ulich, Germany
}%
\author{L.A.~Kondratyuk}
\affiliation{%
Institute of Theoretical and Experimental Physics,
B.~Cheremushkinskaya 25, 117218 Moscow, Russia
}%


\date{\today}% It is always \today, today,

\begin{abstract}
We investigate the $K^-\,^3 \mathrm{He}$ and $K^+ K^-$ interactions
in the reaction $pd \to {}^3 \mathrm{He} K^+ K^- $ near threshold
and compare our model calculations with data from the MOMO
experiment at COSY-J\"ulich. A large attractive effective $K^- p$
amplitude would give a significant $K^-\,{}^3{\mathrm{He}}$
final-state interaction effect which is not supported by the
experimental data. We also estimate upper limits for the
$a_0(980)$ and $f_0(980)$ contributions to the produced $K^+ K^-$
pairs.
\end{abstract}



\maketitle
\section{\label{intro} Introduction}

Low energy $\bar K N$ and $\bar K A$ interactions have been subject of
extensive studies during the last two decades. The well known
phenomenological analysis of $\bar K N$ scattering lengths by
Martin~\cite{Martin81} demonstrated that the $s$-wave $K^-p$
scattering length is large and repulsive, $\mathrm{Re}\, a (K^- p) =
-0.67$ fm, while for the $K^- n$ case it is moderately attractive,
$\mathrm{Re}\,a (K^- n)=0.37$~fm. Recently, new data on the
strong-interaction $1s$ level shift of kaonic hydrogen atoms were
obtained at KEK (KpX experiment)~\cite{Ito,Iwasaki} and Frascati
(DEAR)~\cite{Guaraldo1}. They correspond to the following repulsive
values of the $K^-p$ scattering length
\begin{equation}
   a(K^-p)=-(0.78\pm 0.18)+i(0.49\pm 0.37) ~{\mbox{fm}}
 \label{Iwasaki}
\end{equation}
for KpX, and
\begin{eqnarray}
  a(K^-p)&=&(-0.468 \pm 0.090_{\mbox{stat}}\pm
  0.015_{\mbox{syst}}) \nonumber \\
  & +&i(0.302 \pm
  0.135_{\mbox{stat}}\pm
  0.036_{\mbox{syst}})~{\mbox{fm}}
\label{Guaraldo1}
\end{eqnarray}
for DEAR.

Nevertheless, as it was argued in Refs.~\cite{Dalitz,Weise1}, the
actual $K^-p$ interaction can be attractive if the isoscalar
$\Lambda(1405)$ resonance is a bound state of the ${\bar K}N$ system.
Such a scenario can be explained within Chiral Perturbation Theory
where the leading order term in the chiral expansion for the $K^-N$
amplitude is attractive. Further developments in the analysis of the
${\bar K}N$ interaction based on chiral Lagrangians can be found in
Refs.~\cite{Weise,Oset,Oller01,LuKo,Oller05}. Such a peculiar
behavior of the $\bar K N $ dynamics leads to very interesting
in-medium effects for anti-kaons in finite nuclei as well as in dense
nuclear matter, including neutron stars, see {\em e.g.\/}
papers~\cite{Sibirtsev98,Lutz,Sibirtsev00,Ramos,Heiselberg,Cieply} and
references therein.

Exotic few-body nuclear systems involving the $\bar K$-meson as a
constituent were predicted by Akaishi and Yamazaki \cite{Akaishi02}.
They argued that the ${\bar K}N$ interaction is characterized by a
strong $I{=}0$ attraction, which allows the few-body systems to form
dense and deeply bound $\bar K$-nuclear states.

Evidence for a strange tribaryon $S^0(3115)$ with a width below 21 MeV
was observed in the interaction of stopped $K^-$-mesons with
$^4$He~\cite{Suzuki04}.  This state was interpreted as a candidate for
a deeply bound state $({\bar K}NNN)^{Z{=}0}$ with $I{=}1,I_3{=}{-}1$
\cite{Suzuki04,Akaishi05}. However, the $S^0(3115)$ is about 100 MeV
below the predicted mass, and in the experiment an isospin-1 state was
detected at a position where no such peak was predicted. It was
discussed in Ref.~\cite{Akaishi05} that this discrepancy can be
resolved by tuning parameters of the model~\cite{Akaishi02}.  The
results of Akaishi and Yamazaki~\cite{Akaishi02} were criticized by
Oset and Toki~\cite{OsetToki} who argued that the model of
Ref.~\cite{Akaishi02} is unrealistic. Oset and Toki also suggested
that the peaks in the reaction with ${}^4\mathrm{He}$ can be due to
$K^-$ absorption on a pair of nucleons. This suggestion puts doubt
whether a narrow tribaryon $S^0(3115)$ really exists.

Another tentative evidence for a $K^-pp$ bound state produced in $K^-$
absorption at rest on different nuclear targets was found by the
FINUDA collaboration~\cite{Finuda}. It was detected through its
two-body decay into a $\Lambda$ and a proton. The signal in the
$\Lambda p$ invariant-mass distribution is about 115 MeV below the
expected mass of the $\Lambda p$ system in case of non-bound $K^-NN$
absorption.  Magas {\em et al.\/}~\cite{Magas06} showed that the
FINUDA signal can also be explained by a $K^- pp \to \Lambda p$
reaction followed by final-state interactions (FSI) of the produced
particles with the remnant nucleus.

Thus, it is obvious that further searches for bound kaonic nuclear
states as well as new data on the interactions of $\bar K$-mesons with
light nuclei are of great interest.

In a recent paper~\cite{Grishina05} we presented a first calculation
of the $s$-wave $K^-\alpha$ scattering length $A(K^-\alpha)$ and
discussed how to determine it from the $K^-\alpha$ invariant-mass
distribution in the reaction $dd{\to}\alpha\,K^+ K^-$ near threshold.
In the present paper we consider the $K^-\,^3\mathrm{He}$ FSI in the
reaction $pd \to {}^3 \mathrm{He}\,K^+ K^- $ and compare our
calculations with the existing data on this reaction near
threshold~\cite{Diss_MOMO,MOMO_PRC}.  We also analyze the $K^+ K^-$
relative-energy distribution for this reaction and estimate possible
contributions from the $a_0(980)$ and $f_0(980)$ resonances.

Our paper is organized as follows: In Sect.~\ref{sec:1} we calculate
the $K^-\,^3$He and $K^- \alpha $ scattering lengths.  In
Sect.~\ref{sec:2} an analysis of the $K^-\,^3$He FSI in the reaction
$pd \to {}^3\mathrm{He}\,K^+ K^-$ is presented.  In Sect.~\ref{sec:3}
we analyze the differential $K^+ K^-$ distributions and discuss the
possible contributions from the $a_0/f_0(980)\to K^+K^-$ channels.
Our conclusions are given in Sect.~\ref{sec:concl}.

\section{\label{sec:1}  \boldmath $K^-\,^3\mathrm{He}$ and $K^-\alpha$ scattering lengths}

In order to calculate the $s$-wave $K^-\,^3$He and $K^-\alpha$
scattering lengths and corresponding enhancement factors we use the
multiple-scattering approach (MSA) in the fixed center approximation
described in detail in our previous paper~\cite{Grishina05}. For the nuclear
density we use a factorized model with the single-nucleon density in
Gaussian form
\begin{eqnarray}
  && \rho (\mathbf{r})= \frac{1}{(\pi \, R^2)^{3/2}} \
  \mathrm{e}^{-r^2/R^2} \ , \label{radiushe4}
\end{eqnarray}
where $R^2/4=0.62$ and 0.7~fm${}^{2}$ for $^{4}\mathrm{He}$ and
$^{3}\mathrm{He}$, respectively. Note, that the independent particle
model gives a rather good description of the $^{4}\mathrm{He}$ and
$^{3}\mathrm{He}$ electromagnetic form factors up to a momentum
transfer $\mathbf{q}^2=8$~fm${}^{-2}$ (see {\em e.g.\/}
Ref.~\cite{Boitsov}).

Some theoretical predictions for the $K^- \alpha$ and
$K^-\,^{3}{\mathrm{He}}$ scattering lengths, $A(K^- \alpha)$ and
$A(K^-\, ^{3}{\mathrm{He}})$, have been published in
Refs.~\cite{Grishina05,Kondrat05}. In Table~\ref{Tab3} we present new
results calculated for different $\bar K N$ inputs as compared to
Refs.~\cite{Grishina05,Kondrat05}.  We consider the $\bar K N$
scattering lengths from a $K$-matrix fit (Set~1)~\cite{Barret} as well
as the predictions for the ${\bar K}N$ scattering amplitudes based on
the chiral unitary approach of Ref.~\cite{Oller01} (Set~2).  The
constant scattering-length fit from Conboy~\cite{Conboy} is denoted as
Set~3. We note that the ${\bar K}N$ scattering lengths described by
Sets ~1--3 correspond to their vacuum values.  At the same time
Sets~4--5 describe the effective ${\bar K}N$ scattering lengths that
contain in-medium effects.

One of the most extensive analyses of the effective $\bar K N$
interactions in nuclear medium has been presented by Ramos and
Oset~\cite{Ramos00} within a self-consistent microscopic theory.  The
resulting $K^-$ attraction in medium has been found to be smaller than
predicted by other theories and approximation schemes. The
isospin-averaged effective $\bar K N$ scattering length is moderately
attractive and its real part does not exceed the value of
\begin{equation}
  \mathrm{Re} \ a^{\mathrm{eff}} \simeq 0.3 \, \mbox{fm} \ ,
\label{Ramos}
\end{equation}
at nuclear density $\rho \geq 0.3 \rho_0$.  The obtained shallow
$K^-$-nucleus optical potential with a depth of $-50$~MeV (for the
real part of the potential at $\rho=\rho_0$) was successfully used to
reproduce the experimental shifts and widths of kaonic atoms over the
periodic table~\cite{Hirenzaki00}.

In contrast to the results of Ref.~\cite{Ramos00}, Akaishi and
Yamazaki~\cite{Akaishi02} proposed much more attractive optical
potential which corresponds to the following effective $\bar K N$
scattering lengths for the $I=0,1$ channels in the nuclear medium
\begin{eqnarray}
  a_0^{\mathrm{eff}}&=&2.9+i1.1 \, \mbox{fm} \ , \nonumber \\
  a_1^{\mathrm{eff}}&=&0.43+i 0.30 \, \mbox{fm} \ .   \label{AY}
\end{eqnarray}
According to the Akaishi and Yamazaki approach, such a strong
attraction appears already in the case of few-nucleon systems
generating deeply bound $\bar{K}$-nuclear states~\cite{Akaishi02}.

In order to demonstrate the sensitivity of our results to possible
modifications of the $\bar K N$ scattering amplitudes in the presence
of nuclei we consider as Set~4 the moderately attractive effective
scattering length from Ref.~\cite{Ramos00}. As Set~5 we choose the
strongly attractive in-medium solution found in
Refs.~\cite{Akaishi02,Akaishi05} and given by Eq.(\ref{AY}).

The calculated values of the $A(K^-\alpha)$ and $A(K^-
\,^3{\mathrm{He}})$ within the multiple-scattering theory are listed
in the 5$^{\mathrm{th}}$ and 6$^{\mathrm{th}}$ columns of
Table~\ref{Tab3}.  They are very similar for Sets 1 and 2,
$A^{\mathrm{MS}}(K^- \alpha) \sim (-1.9 + i 1.0)$~fm and
$A^{\mathrm{MS}}(K^- \,^3{\mathrm{He}}) \sim (- 1.6 + i 1.0)$~fm.  The
results for Set~3 are quite different especially for the imaginary
part of $A^{\mathrm{MS}}(K^-\,^3{\mathrm{He}})$ and the real part of
$A^{\mathrm{MS}}(K^-\alpha)$. The calculations with the effective
$\bar K N$ amplitude from Ref.~\cite{Ramos00} give the $K^- \alpha$
scattering length with an imaginary part roughly two times larger than
the result obtained with the vacuum $\bar K N$ scattering lengths. Not
surprisingly, the exotic Set~5 for the elementary amplitudes extracted
from Refs.~\cite{Akaishi02,Akaishi05} leads to enormously large
scattering lengths for $K^- \alpha$ and $K^-\,^3{\mathrm{He}}$ systems
with real parts of $-3.5$~fm and $-4$~fm, respectively.  In the case
of the Set~5 we also performed calculations using the single-nucleon
density parameters in Eq.~(\ref{radiushe4}) from Ref.~\cite{Akaishi02}
$R^2/4=0.48$~fm${}^{2}$ and 0.64~fm${}^{2}$ for $^{4}\mathrm{He}$ and
$^{3}\mathrm{He}$, respectively.  The corresponding results, presented
in square brackets in Table~\ref{Tab3}, show a not very high
sensitivity to the parameter $R$.

\begin{table*}
  \caption{$K^- \alpha$ and $K^-\,^3$He scattering lengths obtained
    within the multiple-scattering approach and $K^- \alpha$ scattering
    length calculated using the optical-potential model for
    various choices of the elementary $\bar K N $ scattering lengths
    $a_I({\bar{K}N})$ ($I=0,1$). The effective $\bar K N$ scattering length
    of Set~4 is extracted from the isospin averaged optical potential.
    Therefore it can only be applied for the calculation of $A(K^- \alpha)$.}
   \label{Tab3}
   \begin{ruledtabular}
   \begin{tabular}{lclllll}
  Set & Ref. & $\vphantom{\frac{A^{I}}{F}} a_0(\bar{K}N)\ [{\mbox{fm}}]$&
  $a_1(\bar{K}N)\ [{\mbox{fm}}]$& $A^{\mathrm{MS}}(K^- \alpha)\ [{\mbox{fm}}]$
  & $A^{\mathrm{MS}}(K^- {}^3{\mathrm{He}})\ [{\mbox{fm}}]$ &
$A^{\mathrm{opt}}(K^- \alpha)\ [{\mbox{fm}}]$ \\
  \hline
  1 & \cite{Barret} & $-1.59+i0.76$ & $0.26 + i0.57$ & $-1.80+ i0.90$ & $-1.50+i0.83$&
$-1.26+i0.60$\\
  2 & \cite{Oller01} & $-1.31+i1.24$ & $0.26 + i0.66$ & $-1.98+ i 1.08$ & $-1.66 + i1.10$&
$-1.39 + i0.65$\\
  3 & \cite{Conboy} & $-1.03+i0.95$ & $0.94 + i0.72$ & $-2.24+ i 1.58$ & $-1.52 + i1.80$&
$-1.59+i0.88$\\
  4 & \cite{Ramos00} & \multicolumn{2}{c}{$0.33 +i 0.45$  isospin average}
  & $-1.47 + i 2.22$ & &
$-1.51+i1.20$\\
  5 & \cite{Akaishi02} & $2.9+i1.1$ & $0.43 + i0.30$ & $-3.49 + i 1.80$ & $-3.93 + i4.03$&
$-1.57+i0.74$\\
&  &  & & $[-2.99 + i 1.27]$ & $[-3.91 + i3.62]$&
$[-1.31+i0.73]$\\
\end{tabular}
\end{ruledtabular}
\end{table*}

Alternatively we considered the optical potential
\begin{equation}
V^{\mathrm{opt}}_{K^- A}(\mathbf{r})=-\frac{2\, \pi}{\mu _{\bar{K}N}} \
\left[ a_{K^- p}\, Z\rho_p(r) + a_{K^- n} N \rho_n(r)\right] \ ,
\label{Opt}
\end{equation}
where $\mu _{\bar{K}N}$ is the reduced mass of the $\bar{K}N$ system
and $\rho_p(r)=\rho_n(r)=\rho(r)$ is defined by Eq.~(\ref{radiushe4})
with eliminated c.m. motion, i.e.~$R^2 \to R^2 \ (A-1)/A$.  The
results for $A(K^- \alpha)$ are presented in the last column of the
Table~\ref{Tab3}. $A^{\mathrm{opt}}(K^- \alpha)$ was found to be about
30--40\% smaller than $A^{\mathrm{MS}}(K^- \alpha)$ for the vacuum
parameters of the $\bar{K} N$ interactions.  In the case of Set~5 the
$K^- \alpha $ scattering length was calculated using the
optical-potential model with $R^2/4=0.62$~fm${}^{2}$ or
$R^2/4=0.48$~fm${}^{2}$ (the latter solution is presented in square
brackets).  This result is more than a factor of two smaller as
compared to the solution for Set~5 obtained in the framework of the
MSA.

The single-scattering term of the total $K^- A$ scattering length is
the same in both the muliple-scattering and optical-potential schemes.
It is equal to $$
A^{(1)}(K^- A)=\frac{\mu_{\bar{K}
    A}}{\mu_{\bar{K}N}} \ (Na_{\bar{K}n}+Za_{\bar{K}p}) \ , $$
where
$\mu_{\bar{K} A}$ is the $K^- A$ reduced mass.  Note that in the case
of the $K^-$d system the higher order rescattering corrections were
analyzed within the fixed center approximation to the Faddeev
equations with the input parameters from the chiral unitary
approach~\cite{Kamalov01}.  Their contributions were evaluated and
found to be very noticeable.  Within the optical-potential model the
next-to-leading order scattering terms $A^{(n)}(K^- A)$ are defined
using different averaging procedures over the coordinates of the
nucleons than in the case of the MSA in the fixed center
approximation. However the total meson-nucleus scattering length
obtained within the simplest optical-potential model needs some
corrections especially important for the few-body systems.
Particularly, in Ref.~\cite{Wycech95} the isospin-symmetric case of
the $\eta$-nucleus system was considered with equal elementary
scattering amplitudes of the $\eta$ meson on the proton and neutron.
For the $\eta$-nucleus scattering length calculated using the
optical-potential model it was suggested that each multiple-scattering
term of $n$-th order should be corrected with q factor $((A-1)/A)^n$
to remove multiple collisions on the same nucleon.  Such a correction
cannot be used for the $K^- A$ optical-potential calculations due to
the significantly different $K^- n$ and $K^- p$ scattering lengths.
The advantage of the MSA applied to the $K^- A$ system is that it
explicitly takes into account the difference between the elementary
$K^- p$ and $K^- n$ scattering amplitudes~(see
Ref.~\cite{Grishina05}).  Furthermore, if the $\bar K N$ interaction
radius is small as compared to the size of the nucleus, the MSA is
valid and the overall scattering amplitude can be expressed in terms
of the individual on-shell meson-nucleon amplitudes without extra free
parameters. The $\bar {K} N$ potential constructed by Akaishi and
Yamazaki is of short range type with $R=0.66$~fm (see
Ref.~\cite{Akaishi02}). Using this assumption we can reliably apply
the MSA for the description of the $K^-$-light nucleus system.

\section{\label{sec:2} \boldmath $K^-\,^3\mathrm{He}$ FSI in
$pd{\to}{}^3\mathrm{He}\,K^+ K^-$}

\begin{figure}
\begin{center}
  \scalebox{0.95}[0.95]{\resizebox{\columnwidth}{!}{
  \includegraphics{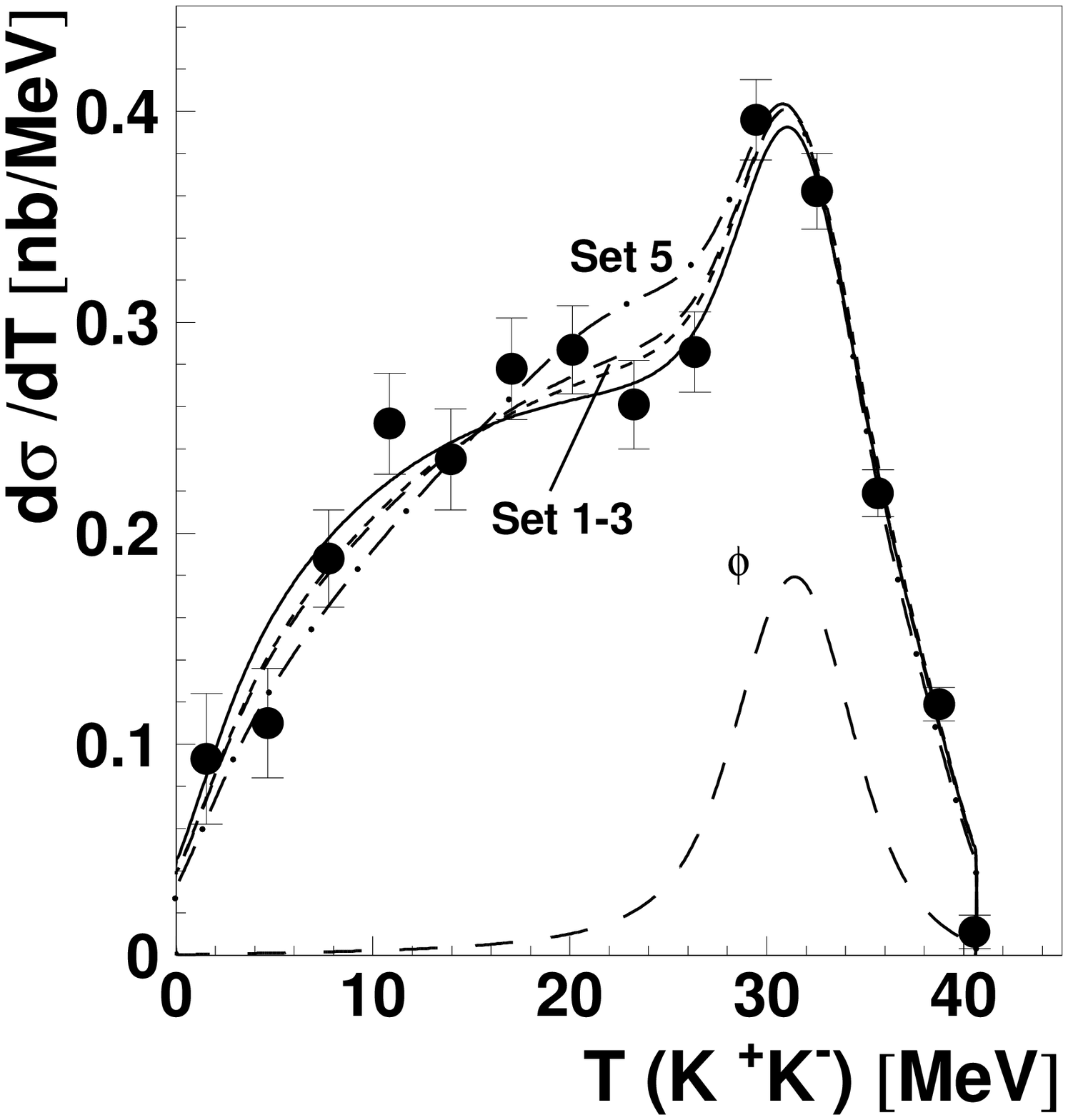}}}
  \scalebox{0.95}[0.95]{\resizebox{\columnwidth}{!}{
  \includegraphics{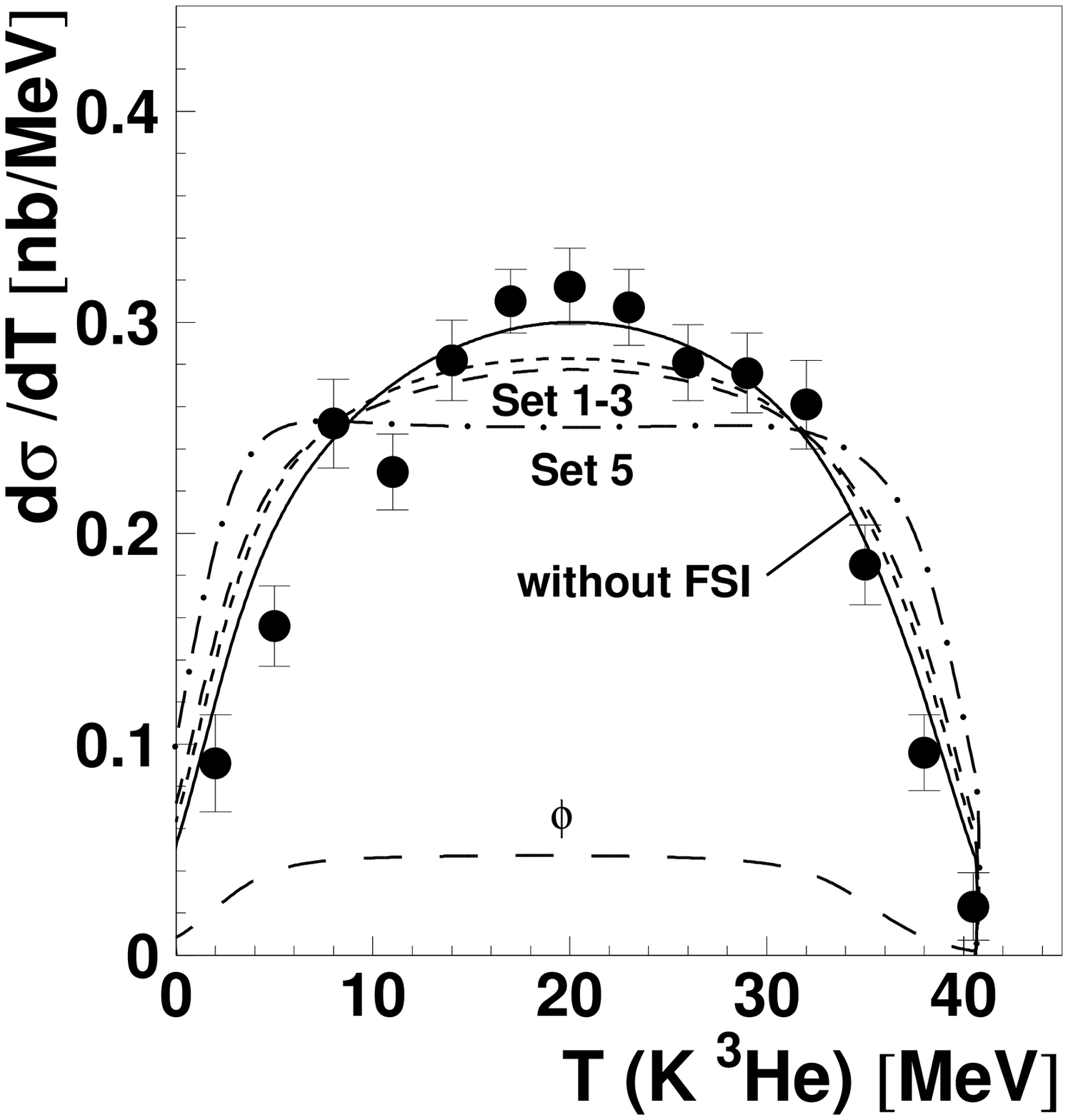}}}
  \vspace*{-2mm}
  \caption{\label{fig:Khe} Distribution of the $K^+ K^-$ (upper) and
    $K\, {}^3\mathrm{He}$ (lower) relative energies for the $pd\to{}^3
    \mathrm{He}\,{K^+ K^-}$ reaction at an excess energy of 41~MeV.
    The MOMO data are taken from Refs.~\cite{Diss_MOMO,MOMO_PRC}. The
    solid line describes the incoherent sum of a pure phase-space
    distribution and the $\phi(1020)$ contribution (long-dashed line).
    The short-dashed and dashed lines show the effect of the $K^- \,
    {}^3 \mathrm{He}$ FSI for parameters of Set~1 and~3, respectively.
    The dash-dotted line shows the effect for the strongly modified
    $\bar {K} N$ scattering lengths in nuclear medium~\cite{Akaishi02}
    leading to the deeply bound states.}
\end{center}
\end{figure}

We now discuss the $K^-\, ^3{\mathrm{He}}$ FSI effect in the reaction
$pd \to {}^3{\mathrm{He}}\, K^+K^-$ near threshold and compare our
calculations performed in the MSA with the data from the MOMO
experiment \cite{Diss_MOMO,MOMO_PRC} at COSY-J\"ulich.  The MOMO
collaboration measured at three different beam energies, corresponding
to excess energies of $Q=35,\ 41$ and 55 MeV with respect to the
$K^+K^-$ threshold.  We only consider the data at the central energy
since these constitute the best compromise between available phase
space and resolution for our analyses.  The MOMO data are presented in
terms of relative energies $T_{K^- \,^3\mathrm{He}}$ and $T_{K^+
  K^-}$; in the non-relativistic approximation they can be expressed
through the corresponding invariant masses $M_{ij}$ by
$T_{ij}=M_{ij}-m_i-m_j$.  Since the MOMO experiment was not sensitive
to the charge of the detected kaons, the measured
$T(K,\,^3\mathrm{He})$ distributions (see Fig.~\ref{fig:Khe}) are
symmetric with respect to $Q/2$. This is taken into account in our
calculations by constructing the half-sum of the $K^+$ and $K^-$
contributions.

As the first step of our analysis we neglect all FSI effects and
investigate the contribution of the $\phi(1020)$ meson by fitting the
$K^+ K^-$ relative-energy distribution (taking into account the
experimental mass resolution quoted in Ref.~\cite{MOMO_PRC}). The
$\phi$-meson contribution is found to be about 16\% of the total cross
section ($9.6\pm 1.0$)~nb, which is in agreement with the result from
Ref.~\cite{MOMO_PRC}.  In Fig.~\ref{fig:Khe} we show the $K^+K^-$
relative-energy distribution; the solid line describes the incoherent
sum of a pure phase-space distribution and the $\phi(1020)$
contribution.

The short-dashed and dashed lines in Fig.~\ref{fig:Khe} show the
influence of the $K^{-}\,^3 \mathrm{He}$ FSI on the $K^+K^-$
relative-energy distribution.  The dash-dotted line shows the effect
for the strongly modified $\bar {K} N$ scattering lengths in nuclear
medium (Set 5) leading to the deeply bound states.

In Fig.~\ref{fig:Khe} we also present calculations of the
$K\,^3{\mathrm{He}}$ relative-energy spectrum.  The predictions are
normalized to the total $p d \to {}^3{\mathrm{He}}\,K^+K^-$ cross
section of 9.6~nb.  The dash-dotted line, corresponding to Set~5,
demonstrates a pronounced deformation of the $K\, ^3 \mathrm{He}$
relative-energy spectrum in the region of small energies. It is in
clear contradiction to the data.

While the solution without FSI is in best agreement with the data, the
results with $K^-\,^3\mathrm{He}$ FSI calculated using elementary
$\bar{K} N$ amplitudes from Sets 1--3 cannot be ruled out due to the
uncertainties of the MSA (see {\em e.g.\/} Ref.~\cite{Grishina05}) and
the experimental errors.

\section{\label{sec:3} \boldmath $K^+ K^-$ energy spectrum and
  $a_0(980)/f_0(980)$ production}

The $a_0$ and $f_0$ resonances may give some contributions to the
$pd \to {}^3\mathrm{He}\,K^+ K^-$ cross section. In this case one can
write the invariant $K^+ K^-$ mass distribution as
\begin{equation}
  \frac{\mathrm{d} \sigma_{pd \to {}^3\mathrm{He} \, K^+ K^-}}{\mathrm{d}M}=
  \frac{\mathrm{d} \sigma_{\mathrm{BG}}}{\mathrm{d}M} +
  \frac{\mathrm{d} \sigma_{\phi}}{\mathrm{d}M}+
  \frac{\mathrm{d} \sigma_{a_0}}{\mathrm{d}M}+
  \frac{\mathrm{d} \sigma_{f_0}}{\mathrm{d}M}\  .
\label{eq:dsdmtot}
\end{equation}

The first term describes the non-resonant $K^+ K^-$ production with a
constant interaction amplitude near threshold. The $K^- \,
^3\mathrm{He}$ FSI effects can be neglected since their influence on
the $K^+ K^-$ distribution is very small, see Fig.~\ref{fig:Khe}.  The
$\phi (1020)$-meson contribution $\mathrm{d}
\sigma_{\phi}/\mathrm{d}M$ has already been considered in the previous
section. The last two terms reflect the contributions from the
$a_0(980)$ and $f_0(980)$ resonances.  Each of them can be written as
a product of the total $a_0$- or $f_0$-production cross section
$\sigma_{a_0}$ ($\sigma_{f_0}$) as a function of the ``running'' mass
$M$ and the Flatt\'e mass distribution.  For example, in case of $a_0$
production we have
\begin{eqnarray}
  &&\frac{\mathrm{d}\sigma _{a_0 K^+ K^-}}{\mathrm{d} M^2} (s,M) = \sigma_{a_0}(s,M) \times
  \nonumber \\
  &&\ C_{\mathrm{F}}
  \frac{M_R \Gamma_{a_0 K^+ K^-}(M)} {(M^2-M_R^2)^2 + M_R^2
    \Gamma_{\rm tot}^2(M)} \label{dsdmKK}
\end{eqnarray}
with the total width $\Gamma_{\rm tot}(M)=\Gamma_{a_0 K\bar K}(M)+
\Gamma_{a_0 \pi\eta}(M)$ and $\Gamma_{a_0 K^+ K^-}(M)=0.5\,
\Gamma_{a_0 K \bar K}$. The constant $C_{\mathrm{F}}$ is introduced to
normalize the total decay probability of the $a_0$ to unity. The
partial widths
\begin{eqnarray}
  \Gamma_{a_0 K\bar K}(M) &=& g_{a_0 K\bar K}^2 \frac{q_{K\bar K}}
    {8\pi M^2}, \nonumber\\
  \Gamma_{a_0 \pi\eta}(M) &=& g_{a_0
    \pi\eta}^2 \frac{q_{\pi\eta}}{8\pi M^2}
 \label{width}
\end{eqnarray}
are proportional to the decay momenta in the c.m.\/ system
\begin{eqnarray}
  q_{K\bar K}\! &=& \frac{\left[(M^2-(m_{K}+m_{\bar K})^2)
      (M^2-(m_{K}-m_{\bar K})^2)\right]^{1/2}}{2M} \nonumber\\
  q_{\pi\eta}&=&\frac{\left[(M^2-(m_{\pi}+m_{\eta})^2)
      (M^2-(m_{\pi}-m_{\eta})^2)\right]^{1/2}}{2M}\ ,
  \nonumber
\end{eqnarray}
for a particle of mass $M$ decaying to $K\bar K$ and $\pi\eta$,
respectively. The contribution of the $f_0$-meson can be written in a
similar manner taking into account its decays into $\pi \pi$ and
$K\bar{K}$. The parameters $g_{a_0 \pi \eta}$, $R_{a_0}=g_{a_0
  K\bar{K}}^2/g_{a_0 \pi\eta}^2$, $M_R$ and $g_{f_0 \pi \pi}$,
$R_{f_0}=g_{f_0 K\bar{K}}^2/g_{f_0 \pi\pi}^2$, $M_R$ of the Flatt\'e
amplitudes for the $a_0$ and $f_0$ resonances can be taken from
literature (see {\em e.g.\/} most recent
papers~\cite{Achasov03,Anisovich04,Baru04} and references therein).

Using Eq.~(\ref{eq:dsdmtot}) we calculate the $K^+ K^-$ mass
distributions with parameters of Set $a_0$[Crystal
Barrel]~\cite{Abele98} and Set $a_0$[E852]~\cite{Teige01} for the
$a_0(980)$ resonance contribution as well as Set
$f_0$[BES]\cite{Ablikim05} and Set $f_0$[E791]\cite{Aitala01} for the
$f_0(980)$.  These parameters are presented in Table~\ref{Tab2}.

\begin{table}
  \caption{Flatt\'e parameters for the $a_0(980)$ and $f_0(980)$
    resonances.}
  \label{Tab2}
  \begin{ruledtabular}
  \begin{tabular}{lcccc}
           Set   & Ref. &$g_{a_0 \pi \eta}$ or $g_{f_0 \pi \pi}$      &
               $R_{a_0}$ or $R_{f_0}$ &$M_{R}$       \\
                 &      &                                        [GeV]&
                                               [GeV]\\
           \hline
      $a_0$[CB]& \cite{Abele98}& $2.3$ & $1.03$ & $0.999$ \\
      $a_0$[E852] & \cite{Teige01}& $2.47$ & $0.91$ & $1.001$ \\
      & & & & \\
      $f_0$[BES] & \cite{Ablikim05}&$1.17$ & $17.72$ & $0.965$ \\
      $f_0$[E791]& \cite{Aitala01} &$1.04$ & $0.22$ & $0.977$ \\
  \end{tabular}
  \end{ruledtabular}
\end{table}

\begin{figure}
\begin{center}
  \scalebox{0.9}[0.85]{
  \resizebox{\columnwidth}{!}{
  \includegraphics{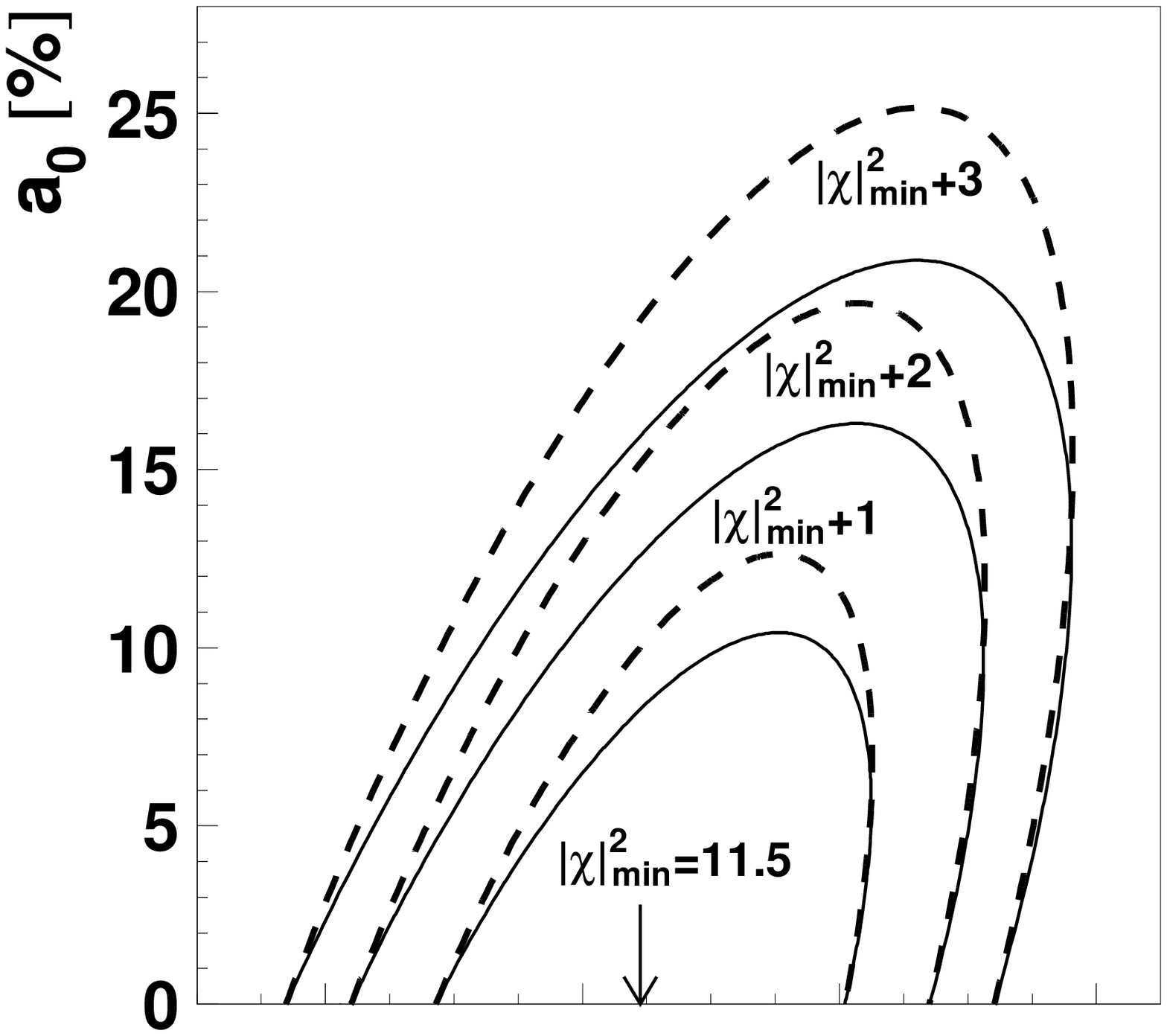}}}
  \scalebox{0.9}[0.85]{
  \resizebox{\columnwidth}{!}{
  \includegraphics{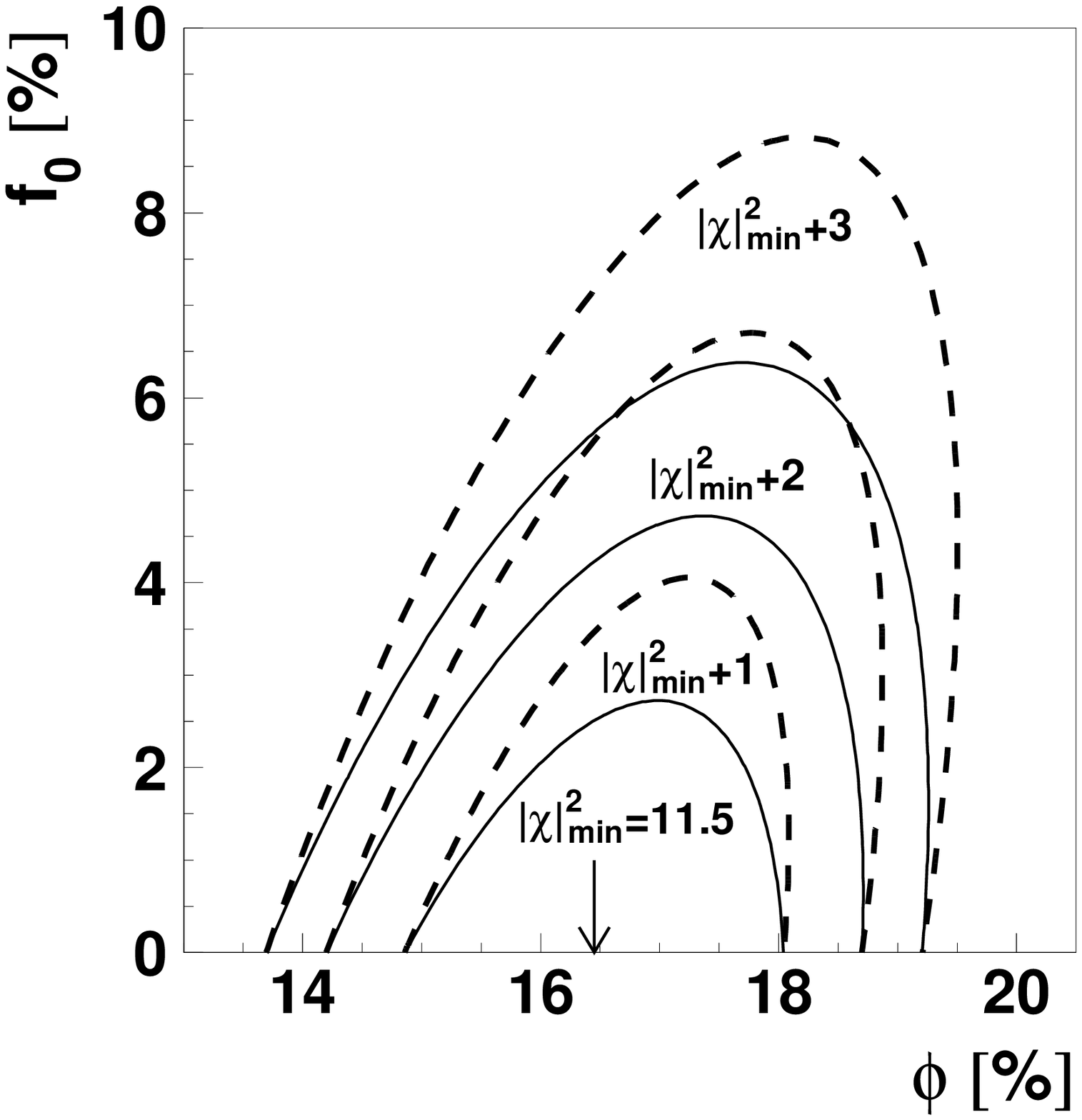}}}
  \vspace*{-5mm}
  \caption{\label{fig:chi2} Result of our fit to the experimental $K^+
    K^-$ mass distribution for the $pd\to{}^3 \mathrm{He}\,{K^+ K^-}$
    reaction at an excess energy of 41~MeV. The $\chi ^2=
    \chi_{\mathrm{min}}^2+1, \chi_{\mathrm{min}}^2+2,
    \chi_{\mathrm{min}}^2+3$ contour lines are obtained for relative
    contributions of $a_0(980)$ or $f_0(980)$ as a function of the
    $\phi(1020)$-meson fraction.  In the upper figure the solid and
    dashed lines were calculated using the Flatt\'e distributions for
    the $a_0$ meson with the parameters of Set $a_0$[Crystal Barrel]
    and Set $a_0$[E852], respectively. In the lower figure the solid
    and dashed contour lines correspond to the contribution of the
    $f_0$ meson with the Flatt\'e parameters of Set $f_0$[BES] and Set
    $f_0$[E791].}
\end{center}
\end{figure}

We then compare the shape of the calculated spectra with that of the
measured $T(K^+K^-)$ distribution.  The solution without $a_0$ and
$f_0$ resonances is in best agreement with the data with
$\chi^2_{\mathrm{min}}=11.5$. The different curves in
Fig.~\ref{fig:chi2} represent the relative contributions of the
$a_0(980)$ or $f_0(980)$ meson versus the fraction of the $\phi(1020)$
meson obtained at $\chi^2=\chi_{\mathrm{min}}^2+1,
\chi_{\mathrm{min}}^2+2$ and $\chi_{\mathrm{min}}^2+3$.  It is seen
that the $a_0(980)$ contribution might reach 20--25\% within a
$\chi_{\mathrm{min}}^2+3$ limit while that from the $f_0$ does not
exceed $\sim10$\% of the total $pd \to {}^3 \mathrm{He}\, K^+ K^-$
cross section at $Q=41$~MeV.

To describe the MOMO data we just added the differential cross
sections for the various channels, neglecting possible interference
between the $a_0$-, $f_0$- and non-resonant contributions to the full
$pd \to {}^3\mathrm{He}\, K^+ K^-$ amplitude. There is no simple way
to calculate interference terms that depend on the spin structure and
relative phases of different amplitudes. Therefore, using the $K^+
K^-$ relative-energy spectrum one can only obtain qualitative
estimates of the $a_0$ and $f_0$ resonance contributions. Nevertheless
we conclude that in the $pd \to {}^3\mathrm{He}\, K^+ K^-$ reaction
near the threshold the $K^+ K^-$ pairs are mainly produced
non-resonantly.

\section{\label{sec:concl} Conclusions}

We present predictions for the $K^-\,^3$He and $K^- \alpha$ scattering
lengths obtained within the framework of the multiple-scattering
approach. We have studied uncertainties of the calculations due to the
presently available elementary $\bar{K} N$ scattering lengths. We have
compared the results for $A(K^- \alpha)$ with values obtained using
the optical-potential model. We have also considered the $K^- \,
{}^3\mathrm{He}$ and $K^+ K^-$ final-state interactions in the
reaction $pd \to{}^3\mathrm{He}\,K^+ K^-$ near threshold and compare
our model calculations with the existing data from the MOMO
collaboration~\cite{Diss_MOMO,MOMO_PRC}. We find that a strongly
modified $\bar {K} N$-effective scattering length extracted from the
very attractive $K^- A$ potential proposed by Akaishi and Yamazaki
\cite{Akaishi02} might lead to a pronounced deformations of the
$K^-\,^3$He and $K^+ K^-$ relative-energy spectra which are in
contradiction to the data. We also derive upper limits on the
$a_0(980)$- and $f_0(980)$-production rates and find them to be on a
level of about 25\% for the $a_0$ and 10\% for the $f_0$ at an excess
energy of 41~MeV.

\section{Acknowledgments}
We are grateful to Hartmut Machner, Hans Str\"oher and Colin Wilkin
for fruitful discussions.  This work was supported by DFG grant 436
RUS 113/787 and RFBR grant 06-02-04013 and the Russian Federal Agency
of Atomic Energy.  V.G. acknowledges support by the COSY FFE grant
No.\/ 41520739 (COSY-071).

\end{document}